\documentclass[journal]{IEEEtran}

\usepackage[pdftex]{graphicx}
\usepackage{amsmath,amssymb,amsfonts}
\usepackage{amsthm}
\usepackage{diagbox}
\usepackage{makecell}
\usepackage{amsmath}
\usepackage{amstext}
\usepackage{textcomp}
\usepackage{mathtools}
\usepackage{cite}
\usepackage{booktabs}
\usepackage{algorithmic}
\usepackage{algorithm}
\usepackage{array}
\usepackage{url}
\usepackage{multirow}
\usepackage{nomencl}
\usepackage{soul}
\usepackage{graphicx}
\usepackage{threeparttable}
\usepackage{tablefootnote}
\usepackage{makecell}
\newtheorem{theorem}{Theorem}

\DeclareMathOperator*{\argmax}{arg\,max}
\DeclareMathOperator*{\argmin}{arg\,min}

\begin{document}

\title{Adversarial Purification for Data-Driven Power System Event Classifiers with Diffusion Models}

\author{Yuanbin~Cheng,~\IEEEmembership{Student Member,~IEEE,}
        Koji~Yamashita,~\IEEEmembership{Member,~IEEE,}
        Jim~Follum,~\IEEEmembership{Member,~IEEE,}
        Nanpeng~Yu,~\IEEEmembership{Senior Member,~IEEE}
\thanks{Y. Cheng, K. Yamashita, and N. Yu are with the Department of Electrical and Computer Engineering, University of California, Riverside, CA 92521, USA. J. Follum is with the Pacific Northwest National Laboratory, Richland, WA, USA.
Corresponding Author: Nanpeng Yu. Email: nyu@ece.ucr.edu}
}


\maketitle

\begin{abstract}
The global deployment of the phasor measurement units (PMUs) enables real-time monitoring of the power system, which has stimulated considerable research into machine learning-based models for event detection and classification. 
However, recent studies reveal that machine learning-based methods are vulnerable to adversarial attacks, which can fool the event classifiers by adding small perturbations to the raw PMU data. 
To mitigate the threats posed by adversarial attacks, research on defense strategies is urgently needed.
This paper proposes an effective adversarial purification method based on the diffusion model to counter adversarial attacks on the machine learning-based power system event classifier.
The proposed method includes two steps: injecting noise into the PMU data; and utilizing a pre-trained neural network to eliminate the added noise while simultaneously removing perturbations introduced by the adversarial attacks.
The proposed adversarial purification method significantly increases the accuracy of the event classifier under adversarial attacks while satisfying the requirements of real-time operations.
In addition, the theoretical analysis reveals that the proposed diffusion model-based adversarial purification method decreases the distance between the original and compromised PMU data, which reduces the impacts of adversarial attacks.
The empirical results on a large-scale real-world PMU dataset validate the effectiveness and computational efficiency of the proposed adversarial purification method.
\end{abstract}


\IEEEpeerreviewmaketitle

\section{Introduction}

Phasor Measurement Units (PMUs) are sensors specifically designed for monitoring power systems in real time. 
By providing synchronized and high-resolution voltage and current measurements, PMUs have revolutionized the monitoring and control of transmission grids \cite{monti2016phasor}.
Although protective relays effectively detect many power system events, such as voltage and frequency events, they remain ineffective in recognizing certain abnormal events, such as oscillations.
For instance, in January 2019, a forced oscillation event in the U.S. highlighted the urgent need for real-time system monitoring with timely corrective actions \cite{NERCreport}. 
Accurate and timely detection and classification of power system events using PMU data is crucial for maintaining power system stability and security \cite{von2017precision} as they can facilitate system operators to take appropriate corrective control actions. 

With the increasing installation of PMUs in the power system \cite{phadke2018phasor}, data-driven event detection methods \cite{cheng2022}, as well as conventional relay-based fault detection methods, have been developed by researchers in academia \cite{jie2020gsp} and industry \cite{NASPI2020Jun}. 
Following the detection of events, the subsequent power system event classification is crucial for distinguishing between different event types, including voltage, frequency, and oscillation events.

In recent years, significant advancements have been made in the field of data-driven power system event classification research.
Due to the rapid advances in computational capabilities and the exponential growth in available PMU data, machine learning methods have shown remarkable proficiency in identifying a wide variety of power system events, such as voltage, frequency, and oscillation events. 
Several machine learning approaches employ feature extraction to address the adverse effects of PMU data quality issues in \cite{9606225}. These approaches include physic-rule based feature extraction \cite{9915447}, event type pattern-based feature engineering \cite{9780546}, physic-rule based decision tree\cite{9072390}, matrix decomposition \cite{9507358}, and energy similarity measurement \cite{8410467}.
Other methods focus on optimizing neural network structures to achieve and enhance the end-to-end classification, such as the convolutional neural network (CNN) \cite{wang2020eventdl}, spatial pyramid pooling (SPP)-aided CNN \cite{Yuan2020}, generative adversarial networks (GANs) \cite{9361704}, and the enhanced ResNet-50 model utilizing information-loading regularization \cite{shi2020power}.
Researchers also explored the hierarchical approach that integrates multiple models for classifying power system events. For instance, \cite{Pavlovski2021} utilizes a hierarchical CNN model with channel filtering, and \cite{8920121} presents a refined two-level hierarchical CNN-based model.
Nevertheless, data-driven power system event classifiers exhibit a notable vulnerability: the low tolerance to adversarial attacks. 
This highlights a potential vulnerability in the practical deployment of such techniques and underscores the necessity of countermeasures.

False Data Injection Attacks (FDIA) are a notable class of cyber-attacks within power systems, compromising the integrity of power system state estimation while avoiding discovery by the residual-based Bad Data Detection (BDD) system \cite{Liu2011fdj}.
Several detection strategies have been developed to mitigate these attacks, including Generalized Likelihood Ratio Test (GLRT) based anomaly detection \cite{6032057}, short-term state forecasting-based anomaly detection \cite{7313024}, and machine learning-based classifiers \cite{6880823}. 
Most of the current literature on FDIA primarily targets protective relays and grid controllers. However, deep learning-based power system event classification algorithms have not been the targets of FDIA research.

Recent developments in cyber security research have highlighted a new category of cyber-attack known as adversarial attacks, particularly within power systems.
These attacks can fool the machine learning-based models by introducing small perturbations to the data.
Originating in the field of computer vision, \cite{goodfellow2014fgsm} illustrated that adding designed imperceptible perturbations to images could lead the machine learning-based classifier to produce incorrect predictions. 
Further research by \cite{cheng2022adv} demonstrates that adversarial attacks are effective on data-driven power system event classifiers, amplifying concerns about the potential adversarial attacks in the power system.
The realization of these threats underscores the need for robust and highly accurate power system event classifiers; without them, the practical implementation in power system operation, control, and protection remains infeasible.
This paper specifically focuses on the defense strategy against the adversarial attack on power system event classifiers.

Numerous adversarial defense techniques have emerged to counter adversarial attacks, broadly categorized into adversarial training and adversarial purification.
Adversarial training, which re-trains classifiers on compromised samples \cite{madry2019deep}, has demonstrated its efficacy against specific attack types \cite{Zhang2019advtrain} for which they are trained. 
However, it is computationally expensive and typically counters only specific attack types.
On the other hand, adversarial purification focuses on pre-processing inputs to remove adversarial perturbations before classification.
This approach typically employs smoothing techniques \cite{Xu_2018} or generative models \cite{pouya2018defensegan,nie2022diffusion} to purify the compromised data.
Adversarial purification presents two significant benefits when compared to adversarial training: 1) it eliminates the need for classifier re-training, and 2) it defends against unforeseen attacks due to its classifier-independent operation. 
However, adversarial purification is generally less effective than adversarial training, as demonstrated in \cite{Croce020a}. 
This performance gap is primarily due to limitations in the current generative model, such as low-quality sampling and mode collapse in GANs\cite{goodfellow14gan}.

This paper primarily concentrates on the adversarial purification method to mitigate the adversarial attacks aimed at the power system event classifier.
Recently, the diffusion model has emerged as a leading generative model \cite{Ho2020ddpm}, outperforming the GANs in high-quality image generation\cite{diff2021beatgan}.
It operates through two multi-step Markov processes: a forward diffusion process that transforms data into Gaussian distribution, and a backward process that reconstructs the data through a series of denoising steps by a trained noise estimator.
The diffusion model has demonstrated exceptional mode coverage, as evidenced by the high test likelihood \cite{song21diff}. 
Given the high-quality sample generation and mode coverage abilities, the diffusion model holds significant promise in addressing the limitations of existing adversarial purification methods.

In this paper, we propose a novel diffusion model-based adversarial purification algorithm against adversarial attacks on the power system event classifier.
Our proposed approach combines truncated forward and backward processes within the diffusion model to form an adversarial purification process.
Computational efficiency is crucial in executing adversarial purification in the power system to support real-time operations, including state monitoring and event detection.
However, the standard diffusion model does not fulfill the real-time requirement due to the multi-step model inferences.
To address the computational challenges arising from multiple model inferences, we employ Denoising Diffusion Implicit Models (DDIM) \cite{song2022denoising} to obviate the need for intermediate inference stages, leading to a significant reduction in inference time and enable the real-time purification process.
Furthermore, an in-depth theoretical exploration of the $L_2$ distance between the original and compromised data illuminates the underlying mechanisms of our adversarial purification technique.

The main contributions of this paper are listed as follows:
\begin{itemize}
    \item This paper proposes an innovative adversarial purification algorithm designed to counter adversarial attacks on the power system event classifier. 
    The proposed algorithm combines the truncated forward and backward processes of the diffusion model to compose a purification process capable of sanitizing compromised PMU data, thereby enabling the classifier to produce correct predictions.
    \item This paper employs the implicit diffusion sampling schedule to substantially reduce the number of iterations of the purification process, thereby leading to a significant reduction in computational time and enabling real-time PMU data processing.
    \item This paper presents a novel theoretical discovery: during the entire purification process, the $L_2$ distance between the original and the compromised PMU data consistently decreases. This discovery reveals the underlying mechanism of our proposed adversarial purification algorithm.
    \item This paper validates the proposed adversarial purification algorithm using a large-scale real-world PMU dataset, encompassing thousands of events from the U.S. power grid. The numerical study results demonstrate the exceptional accuracy and computation efficiency of the proposed adversarial purification algorithm for PMU data.
\end{itemize}

The rest of the paper is organized as follows. 
Section II provides the notations, problem definition of the adversarial attack and purification, and the overall framework of the proposed diffusion-based purification algorithm.
Section III begins with a concise overview of adversarial attack algorithms and the diffusion model, followed by a detailed exposition and the theoretical analysis of our proposed diffusion model-based adversarial purification algorithm.
Section IV presents the numerical evaluation by utilizing multiple adversarial attack algorithms and the comparison results with state-of-the-art adversarial purification algorithms.
Section V concludes the paper with a summary of findings and research directions.

\section{Problem Formulation and Overall Framework}
This section first introduces key notations used to describe the dataset and power system event classifier. 
Then, the problem formulation of the adversarial attack targeting the classifier and the definition of the adversarial purification against such attacks are presented.
The final subsection presents the framework for the proposed diffusion-based adversarial purification algorithm together with an example of real-world PMU data and data-driven event classification results.

\subsection{Notations}

\subsubsection*{Notation 1} 
A PMU time series sample can be represented as a tensor $\mathbf{x} = [\mathbf{m}_1, \mathbf{m}_2,..., \mathbf{m}_W]$, where $\mathbf{x}$ encapsulates PMU measurements spanning a fixed window length of $W$.
For each discrete time step, $i$ ($1 \leq i \leq W$), the matrix $\mathbf{m_i}$ includes four variables derived from an array of PMUs' measurements. The four variables obtained from each PMU are active power ($P$), reactive power ($Q$), voltage magnitude ($|V|$), and frequency ($F$).

\subsubsection*{Notation 2} 
For each PMU time series sample $\mathbf{x}$, there exists a corresponding event label, $y_i$, denoting the power system event type, which is one-hot encoded.
In this paper, the power system event classifier identifies four unique event types: normal behavior, voltage-related events, frequency-related events, and oscillation events. 
As a result, the one-hot encoded event label vector is of length four.

\subsubsection*{Notation 3} 
The power system PMU and event dataset represented as $\mathcal{D} = \{(\mathbf{x}_1, y_1), (\mathbf{x}_2, y_2), \ldots, (\mathbf{x}_n, y_n)\}$, consists of pairs of PMU time series samples and their corresponding event labels, where $n$ denotes the total number of PMU time series samples in the dataset.

\subsubsection*{Notation 4} 
Let $f(\cdot)$ denote a deep learning-based power system event classifier.
Given an input $\mathbf{x}$, the classifier produces a predicted result vector $\hat{y} = f(\mathbf{x})$, which quantifies the probability for each class.

\begin{figure}[t]
\centering
\includegraphics[width=0.485\textwidth]{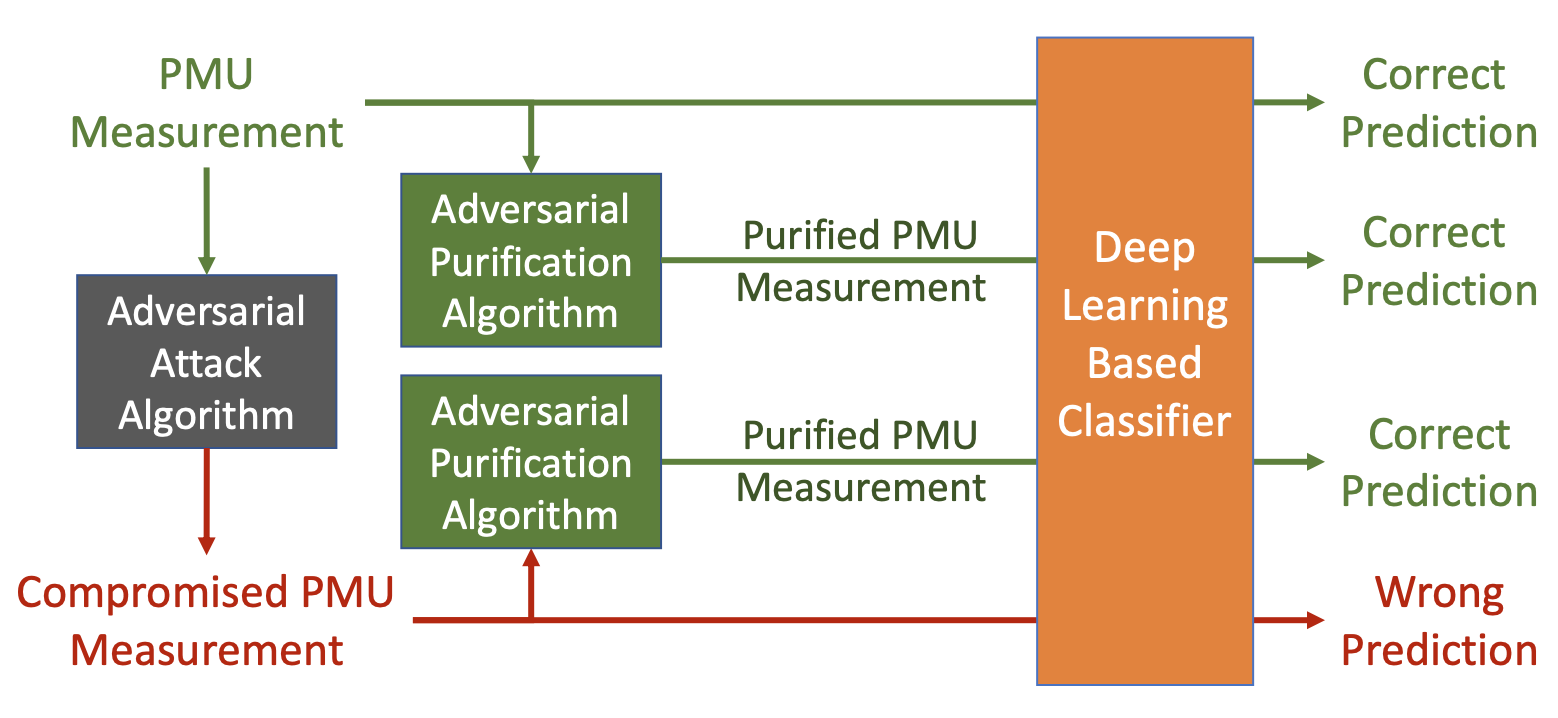}
\DeclareGraphicsExtensions.
\vspace{-1.5em}
\caption{The illustration of adversarial attack and adversarial purification.}
\vspace{-1.5em}
\label{adversarial_attack_purification}
\end{figure}

\vspace{-1em}
\subsection{Problem Definition}

\subsubsection{Power System Event Classification}

The objective of power system event classification is to determine the event type of a PMU time series sample $\mathbf{x}$.
This task can be performed using a deep learning-based classifier, which is trained on the dataset $\mathcal{D}$, including the PMU data and power system event labels. 
The primary goal during the training of the classifier is to minimize the loss function:
\begin{equation}
\argmin_\psi L(f_{\psi}(\mathbf{x}), y),
\end{equation}
where $\psi$ denotes the parameters of the classifier $f_{\psi}(\cdot)$.

In this paper, we employ a state-of-the-art deep neural network-based classifier, specifically an enhanced ResNet-50 \cite{he2015deep}. 
This classifier leverages graph signal processing (GSP) based PMU sorting algorithm and information loading-based regularization to identify power system events. 
Detailed information about the classifier can be found in \cite{shi2020power}.
We primarily focus on this classifier as the target of adversarial attacks and purification throughout this research study.

\subsubsection{Adversarial Attack} 

The adversarial attack aims to discover a perturbation, represented by $\boldsymbol\eta$, for individual samples.
This perturbation is injected on top of the input sample $\mathbf{x}$ to yield the compromised sample, denoted as $\mathbf{x}' = \mathbf{x} + \boldsymbol\eta$.
The primary goal of this attack is to cause the classifier's prediction for $\mathbf{x}'$ to deviate from the true label, $y$, while ensuring the magnitude of $\boldsymbol\eta$ remains sufficiently small to be imperceptible when compared to the original input. 
The problem of identifying the most effective perturbation in the adversarial attack is formulated as:
\begin{equation}
\argmax_\eta L(f(\mathbf{x} + \boldsymbol\eta), y) \text{, subject to } ||\boldsymbol\eta||_2 \leq \xi,
\label{eq:adv_attack}
\end{equation}
where $||.||_2$ denotes the $L_2$ norm, and $\xi$ denotes a small constant that controls the magnitude of the perturbation.

Figure \ref{adversarial_attack_purification} illustrates the procedure of a typical adversarial attack. 
Initially, the deep learning-based classifier correctly classifies the power system event using a PMU time series sample. 
After the adversarial attack is launched on the PMU time series sample, the classifier misclassifies the compromised PMU time series sample, leading to an incorrect event prediction.

\subsubsection{Adversarial Purification}

Adversarial purification is a class of defense algorithms that remove adversarial perturbations with various techniques. 
The primary goal of adversarial purification for power system event monitoring is to correct the misclassified events with compromised PMU samples while maintaining the classification accuracy of samples untouched by adversarial attacks.
We represent the purification process as $g_{\phi}(\cdot)$, which acts on a PMU sample that could be either compromised or uncompromised by adversarial attacks, where the symbol $\phi$ denotes the parameters of the adversarial purification function $g_{\phi}(\cdot)$.
The adversarial purification algorithm manipulates the PMU time series sample to either restore its accurate classification result (in the case of a compromised sample) or preserve its correct classification outcome (for a sample that is not compromised).

Let $\mathcal{A}$ denote the dataset formed by combining the original dataset $\mathcal{D}$, composed of sample-label pairs $(\mathbf{x}, y)$, with a dataset of compromised PMU time series samples consisting of pairs $(\mathbf{x}', y)$. The formal representation of $\mathcal{A}$ is given as:
\begin{equation}
\mathcal{A} = \{(\mathbf{x}_1, y_1), (\mathbf{x}_1', y_1), \ldots, (\mathbf{x}_n, y_n), (\mathbf{x}_n', y_n)\}
\end{equation}

The problem of training the adversarial purification algorithm can be formulated as follows:
\begin{equation}
\argmin_\phi L(f(g_{\phi}(\mathbf{x})), y) \text{,   } (\mathbf{x}, y) \in \mathcal{A}
\end{equation}

Figure \ref{adversarial_attack_purification} also illustrates the adversarial purification process. Through the data processing executed by the adversarial purification algorithm, the correct power system event can be identified for both the original PMU time series samples and the compromised PMU time series samples.

\begin{figure*}[!ht]
\centering
\includegraphics[width=1.0\textwidth]{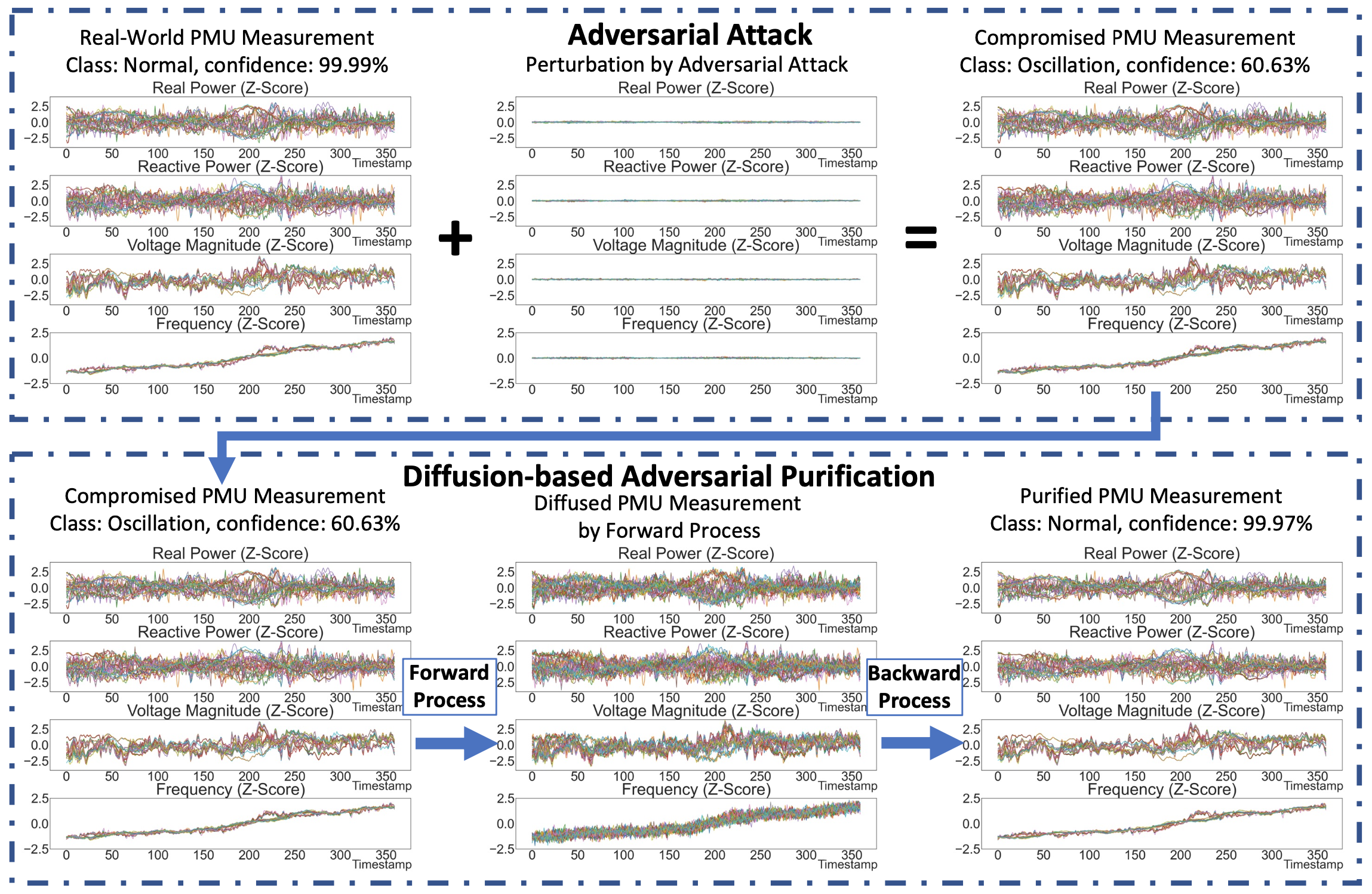}
\DeclareGraphicsExtensions.
\vspace{-1.5em}
\caption{Overall framework of the adversarial attack and the diffusion model-based adversarial purification method on a real-world PMU time series sample.}
\label{OverallFramework}
\vspace{-1.5em}
\end{figure*}

\vspace{-0.5em}
\subsection{Overall Framework}
The overall framework for both the adversarial attack and our proposed diffusion-based adversarial purification algorithm is shown in Fig. \ref{OverallFramework} with a real-world PMU time series sample and event label.

The upper section of  Fig. \ref{OverallFramework} illustrates a successful adversarial attack on a real-world PMU time series sample. 
By injecting a small perturbation, the adversarial attack fooled the event classifier into misclassifying PMU data during normal system operations as an oscillation event.

The lower section of Fig. \ref{OverallFramework} illustrates the procedure of the proposed diffusion model-based adversarial purification method. 
This algorithm comprises two distinct phases: a forward process and a subsequent backward process.
In the forward process, Gaussian noise is incorporated into the data (lower-middle sub-figure).
Conversely, the backward process employs the deep neural network-based noise estimator to effectively eliminate the noise introduced during the forward process (lower-right sub-figure). 
This backward phase not only addresses the Gaussian noise introduced in the forward process but also addresses potential adversarial perturbations injected from adversarial attacks. 
By combining the forward and backward processes, the proposed adversarial purification method can successfully restore the original PMU time series sample, resulting in correct power system event prediction.

\section{Technical Methods}

This section provides the technical methods of the proposed diffusion model-based adversarial purification method.
This section first reviews various adversarial attack algorithms, which later serve as the benchmark in the performance evaluation of adversarial purification methods.
Next, we present the fundamental concepts underpinning the diffusion model, accompanied by the formulations that define the forward and backward processes.
Subsequently, we present our proposed diffusion-based adversarial purification algorithm, providing an in-depth understanding of its internal workings.
In the end, a mathematical analysis offers a theoretical substantiation that our proposed algorithm effectively purifies the perturbations introduced by adversarial attacks.

\vspace{-1em}
\subsection{Review of the Adversarial Attacks}

This section provides a concise overview of five unique adversarial attack algorithms employed in this paper. 
These algorithms were specifically crafted to fool the power system event classifiers, serving as the benchmark for evaluating the effectiveness of adversarial purification algorithms.

\subsubsection{Fast Gradient Sign Method}

Introduced in \cite{goodfellow2014fgsm}, the Fast Gradient Sign Method (FGSM) creates adversarial examples by perturbing the input sample in the direction corresponding to the sign of its gradient. 
Below is the formal representation of this perturbation process:
\begin{equation}
    \boldsymbol\eta = \xi \cdot \text{sign}(\nabla_x J(X, Y_{true})),
\end{equation}
where $\xi$ represents the magnitude of the perturbation, the adversarial event sample $X^{\prime}$ is generated as $X^{\prime} = X + \boldsymbol\eta$. We set $\xi$ = 0.05 in our experiment.

\subsubsection{Basic Iterative Method and Projected Gradient Descent}

The basic iterative method (BIM) \cite{Kurakin2016bim} refines the FGSM approach by introducing iterative perturbation updates. 
Instead of calculating a one-step perturbation, BIM iteratively updates the perturbation over multiple small steps. 
Each small step resembles the FGSM but with a smaller magnitude. 
Additionally, BIM employs clipping to restrict the accumulated perturbations after each step, thus controlling its $L_2$ norm. 
This iterative perturbation technique yields smaller perturbations compared to FGSM.

The Projected Gradient Descent (PGD) \cite{madry2019deep} further enhances the BIM algorithm by initializing the perturbation with a Gaussian distribution instead of zero. 

The BIM and PGD iterative attack methods rely on three critical hyperparameters: 
(1) the number of iterations, denoted as $I$; 
(2) the maximum allowable perturbation, represented as $\xi$; and 
(3) the magnitude of the small perturbation applied per step, denoted as $\alpha$. 
In our experimental setup, we fix these hyperparameters as $I = 100$, $\xi = 0.05$, and $\alpha = 0.005$.

\subsubsection{DeepFool}

The DeepFool algorithm \cite{Moosavi2015deepfool} represents an effective adversarial attack designed to efficiently estimate the minimal norm perturbation required to fool the trained classifier. Utilizing an iterative approach, DeepFool progressively accumulates perturbations until achieving successful misclassification. 
This iterative process comprises two key procedures. 
First, it identifies the nearest decision boundary across all classes except the true label. 
Subsequently, the sample is updated through orthogonal projection onto this identified decision boundary. 
Due to space constraints, this paper omits the implementation details of DeepFool, which are available in \cite{Moosavi2015deepfool}. 
It is noteworthy that the DeepFool algorithm operates without the necessity for hyperparameter tuning.

\subsubsection{Carlini-Wagner L2 attack}

The Carlini-Wagner L2 attack, referred to as the C\&W attack \cite{cw2}, is a powerful adversarial attack method designed to generate small perturbations in input data that can fool the deep learning-based model. 
It formulates the attack as an optimization problem to find the minimum perturbation under $L_2$ norm constraints, maximizing the model's prediction error. 
This method is highly effective and has been widely used to evaluate the performance of adversarial purification algorithms.
We set the number of iterations $I = 50$, learning rate $lr = 0.01$ in the experiment.

\vspace{-1em}
\subsection{Review of the Diffusion Models}

The diffusion models, also known as score-based generative models \cite{Ho2020ddpm, song2021scorebased}, represent a distinct class of generative models.
In contrast to other generative models such as Variational Autoencoders (VAEs) and Generative Adversarial Networks (GANs), diffusion models utilize a Markov chain to progressively denoise a Gaussian noise sample to generate the desired sample.
The training of the diffusion models encompasses two main processes: the forward and the backward processes.

\subsubsection{Forward process}

The forward process starts with a sample, denoted as $\mathbf{x}_0$. In this paper, $\mathbf{x}_0$ is a PMU time series sample.
During the forward process, Gaussian noise is incrementally introduced to $\mathbf{x}_0$ over a series of $T$ steps, guided by a predetermined variance scheduler denoted as $\beta_t$. 
Typically, $\beta_t$ assumes small values within the range of $0.0001$ to $0.02$. 
In this paper, we linearly increase $\beta_t$ within the specified range.
The Gaussian transition kernel employed between each step is formally defined as:
\begin{equation}
q(\mathbf{x}_t|\mathbf{x}_{t-1}) = \mathcal{N}(\mathbf{x}_t; \sqrt{1 - \beta_t} \mathbf{x}_{t-1}, \beta_t \mathbf{I})
\end{equation}

The forward process possesses a remarkable characteristic that enables the direct computation of $x_t$ without the need for iterative methods. 
By introducing the notations $\alpha_t = 1 - \beta_t$ and $\bar{\alpha}_t = \prod_{s=1}^{t} \alpha_s$, the subsequent equation holds:
\begin{equation}
q(\mathbf{x}_t|\mathbf{x}_0) = \mathcal{N}(\mathbf{x}_t; \sqrt{\bar{\alpha}_t} \mathbf{x}_0, (1-\bar{\alpha}_t)I)
\end{equation}

This formulation allows us to directly represent $\mathbf{x}_t$ as a linear combination of $\mathbf{x}_0$ and a Gaussian noise $\boldsymbol\epsilon$.
\begin{equation}
\mathbf{x}_t = \sqrt{\bar{\alpha}_t} \mathbf{x}_0 + \sqrt{(1-\bar{\alpha}_t)} \boldsymbol\epsilon, \text{ } \boldsymbol\epsilon \sim \mathcal{N}(0, I)
\label{eq:forward}
\end{equation}

As $t$ approaches $T$, the value of $\bar{\alpha}_t$ converges towards $0$. Consequently, $\mathbf{x}_t$ approaches a pure standard Gaussian noise, independent of $\mathbf{x}_0$. In the forward process of the diffusion probabilistic models, the original complex data distribution gradually transitions to a simpler one (standard Gaussian distribution) by applying a stochastic transformation with a slowly changing variance.

\subsubsection{Backward process}

The backward process involves denoising the stochastic variable $x_T$ across $T$ sequential steps. 
Similar to the forward process, the backward process can also be characterized at each step by a Gaussian transition kernel, as shown below:
\begin{equation}
p(\mathbf{x}_{t-1}|\mathbf{x}_{t}) = \mathcal{N}(\mathbf{x}_{t-1}; \boldsymbol\mu_\theta(\mathbf{x}_{t}, t), \sigma^2_t \mathbf{I}),
\end{equation}
where $\sigma^2_t$ is the variance, and $\boldsymbol\mu_\theta(\mathbf{x}_{t}, t)$ is the predicted mean of the $\mathbf{x}_{t-1}$ based on $\mathbf{x}_{t}$ and $t$.
The diffusion model employs a trainable noise estimator $\boldsymbol\epsilon_{\theta}(\mathbf{x}_t, t)$, designed to approximate the actual noise, $\boldsymbol\epsilon$, at time $t$ and leverage it for above mean prediction.
The noise estimator is trained to predict the noise introduced at each time step by the forward process, using the following loss function:
\begin{equation}
L(\theta) = \mathbb{E}_q(||\boldsymbol\epsilon - \boldsymbol\epsilon_{\theta}(\mathbf{x}_t, t)||_2),
\label{eq:loss}
\end{equation}
where $\boldsymbol\epsilon$ is a Gaussian noise introduced by the forward process. 
The training of the noise estimator $\boldsymbol\epsilon_{\theta}(\mathbf{x}_t, t)$ takes batches of the inputs $\mathbf{x}_0$ and random timesteps $t$, generates the $\mathbf{x}_t$ by the forward process  \eqref{eq:forward}, then optimize the parameters with respect to the loss function \eqref{eq:loss}. 

Since $x_T$ closely approximates pure Gaussian noise, we can set $\mathbf{x}_T \sim \mathcal{N}(0, I)$ as initialization for the backward process. 
Subsequently, the backward process is capable of generating samples that conform to the input data distribution from initial Gaussian noise, through successive iterative steps.

\subsubsection{Accelerate the backward process}

In the original diffusion model, the backward process denoises over T steps, resulting in slower sample generation compared to other generative models, such as GANs, and VAEs. 
Recognizing that multiple denoising steps can be consolidated into a single non-Markovian step, Song et al. \cite{song2022denoising} proposed the Denoising Diffusion Implicit Model (DDIM), a sampling strategy designed to accelerate the backward process, as shown by:

\vspace{-1em}
\begin{equation}
\begin{split}
\mathbf{x}_{t-1} = (\frac{\mathbf{x}_t - \sqrt{1 - \bar{\alpha}_t} \boldsymbol\epsilon_\theta(\mathbf{x}_t, t)}{\sqrt{{\bar\alpha}_t}}) + \\
\sqrt{1 - \bar{\alpha}_{t-1} - \sigma^2_t} \boldsymbol\epsilon_\theta(\mathbf{x}_t, t) + \sigma_t \boldsymbol\zeta_t,
\end{split}
\label{eq:backward}
\end{equation}
where $ \boldsymbol\zeta_t \sim \mathcal{N}(0, I)$ is standard Gaussian noise independent of $\mathbf{x}_t$, and $\sigma_t$ determines the magnitude of this noise.

In this paper, we incorporate DDIM samplers into our diffusion-based adversarial purification algorithm. 
By setting $\sigma_t = 0$, our diffusion model transforms to an implicit probabilistic model, resulting in deterministic forward and backward processes and removing the effect of the random term.

\vspace{-1em}
\subsection{Diffusion Model-based Adversarial Purification Algorithm for Power System Classifier}

In this subsection, we introduce our algorithm specifically designed for adversarial purification, leveraging the diffusion model.
As discussed in the preceding subsection, the diffusion model is a generative model wherein a trained noise estimator can be employed for sample generation.
The idea of using the diffusion model for real-time adversarial purification of the power system classifier lies in the integration of the truncated forward and backward processes.
Rather than executing the full $T$ steps of the forward process to map the PMU sample to a Gaussian distribution, we truncate both the forward and backward processes at the $T^*$ timestep.
The truncated forward process introduces noise but retains the primary data semantics.
Subsequently, the backward process removes the noise added by the forward process and the potential perturbations induced by cyber-attacks.

\begin{algorithm}[ht]
    \caption{Our Proposed Adversarial Purification Method}
    \label{alg:bim} 
    \hspace*{\algorithmicindent} \textbf{Input:} PMU data sample $\mathbf{x}$, truncated timestep $T^*$, trained noise estimator $\boldsymbol\epsilon_\theta(\mathbf{x_t}, t)$, backward steps $S$\\
    \hspace*{\algorithmicindent} \textbf{Output:} Purified PMU data sample $\mathbf{x}_0$\\
    \vspace{-0.4 cm}
    \begin{algorithmic}[1]
    \STATE \textbf{Forward process (Add noise to sample)} 
    \STATE $\boldsymbol\epsilon \sim \mathcal{N}(0, I)$
    \STATE $\mathbf{x}_{T^*} \leftarrow \sqrt{\bar\alpha_{T^*}}\mathbf{x} + \sqrt{1-\bar\alpha_{T^*}} \boldsymbol\epsilon $
    \STATE \textbf{Backward process (Remove noise from sample)} 
    \STATE $\tau$ $\leftarrow$ [$0$, $\frac{1}{S}T^*$, $\frac{2}{S}T^*$, $\cdots$, $\frac{S-1}{S}T^*$, $T^*$]
    \FOR{i in [$S$, $S-1$, $\cdots$, $1$]}
        \STATE $\mathbf{x}_{\tau_{i-1}} \leftarrow (\frac{\mathbf{x}_{\tau_{i}} - \sqrt{1 - \bar{\alpha}_{\tau_{i}}} \boldsymbol\epsilon_\theta(\mathbf{x}_{\tau_{i}}, {\tau_{i}})}{\sqrt{{\bar\alpha}_{\tau_{i}}}})$
        \STATE $\mathbf{x}_{\tau_{i-1}} \leftarrow \mathbf{x}_{\tau_{i-1}} +
\sqrt{1 - \bar{\alpha}_{\tau_{i-1}}} \boldsymbol\epsilon_\theta(\mathbf{x}_{\tau_{i}}, {\tau_{i}})$ 
    \ENDFOR
    \STATE \textbf{Return the purified PMU data sample $\mathbf{x}_0$} 
    \end{algorithmic} 
    \label{alg:diffusion}
\end{algorithm}

We provide a comprehensive description of the proposed adversarial purification algorithm in \textbf{Algorithm \ref{alg:diffusion}}.
Initially, noise is integrated into the PMU data sample $\mathbf{x}$ (where $\mathbf{x}$ may include the adversarial attack) through a truncated forward process, executing only at step $T^*$.
Consequently, $\mathbf{x}_{T^*}$ is formulated by the amalgamation of noise ($\boldsymbol\epsilon$) and the input sample ($\mathbf{x}$). This amalgamation is controlled by the weights, $\sqrt{\bar\alpha_{T^*}}$ and $\sqrt{1-\bar\alpha_{T^*}}$. 
Subsequently, the backward process removes the incorporated noise and the adversarial perturbation through $S$ total steps, leveraging the trained noise estimator, $\boldsymbol\epsilon_\theta$.
Note that utilizing the DDIM in the backward process enables the skipping of intermediary steps between each iteration.
This reduction in steps not only decreases the number of model inferences but also enhances the computational efficiency of the adversarial purification process. 

The selection of the truncate step, $T^*$, is crucial within the context of our proposed adversarial purification algorithm.
Too much noise has the potential to remove the semantic information (event type) of the original PMU data sample, and too little noise may fail to eliminate the adversarial perturbations. In the proposed algorithm, the total timestep, $T$, involved in the diffusion model, the truncate timestep, $T^*$, and the backward timestep, $S$, are three hyper-parameters that require meticulous tuning to optimize the algorithm’s performance. The detailed methodology and configuration of the hyper-parameters tuning can be found in Section \ref{DataSource}.

In the next subsection, we present a theoretical analysis indicating that the $L_2$ distance between the original and the compromised sample continually diminishes throughout the entire adversarial purification process. 
This continual reduction in $L_2$ distance reveals the fundamental mechanism governing the proposed adversarial purification algorithm.

\vspace{-1em}
\subsection{$L_2$ Distance Analysis of the Diffusion Model Based Adversarial Purification Algorithm}

\subsubsection{$L_2$ Distance Analysis during the Forward Process}

Let $\mathbf{x}$ denote an original PMU data sample, and let $\mathbf{x'}$ represent the corresponding compromised sample produced by an adversarial attack algorithm.
The original and comprised PMU data samples are related to each other via the small perturbation $\boldsymbol\eta$ introduced by the adversarial attack as follows:
\begin{equation}
\mathbf{x'} = \mathbf{x} + \boldsymbol\eta
\end{equation}

The theorem presented below proves that the $L_2$ distance between the perturbed PMU data sample $\mathbf{x'}$ and the original sample $\mathbf{x}$ consistently diminishes throughout the truncated forward process of the adversarial purification algorithm.

\begin{theorem}
The $L_2$ distance between the original PMU data sample $\mathbf{x}$ and the compromised sample $\mathbf{x}'$ decreases after applying the same truncated forward process in the proposed adversarial purification algorithm. Formally, the following inequality holds:
$||\mathbf{x}_{T^*} - \mathbf{x}_{T^*}'||_2 \leq ||\mathbf{x}- \mathbf{x}'||_2$.
\label{theorem:1}
\end{theorem}

\begin{proof}

After undergoing $T^*$ steps of the forward process as shown in \eqref{eq:forward}, the initial PMU data samples $\mathbf{x}$ and $\mathbf{x'}$ evolve into $\mathbf{x}_{T^*}$ and $\mathbf{x}_{T^*}'$, respectively. 
The PMU data samples $\mathbf{x}_{T^*}$ and $\mathbf{x}_{T^*}'$ can be explicitly expressed as follows:
\begin{equation}
\mathbf{x_{T^*}} = \sqrt{\bar{\alpha}_{T^*}} \mathbf{x} + \sqrt{1 - \bar{\alpha}_{T^*}} \boldsymbol\epsilon
\end{equation}
\begin{equation}
\mathbf{x_{T^*}'} = \sqrt{\bar{\alpha}_{T^*}} \mathbf{x}' + \sqrt{1 - \bar{\alpha}_{T^*}} \boldsymbol\epsilon,
\end{equation}
where $\boldsymbol\epsilon$ represents Gaussian noise $\mathcal{N}(0, I)$, which is independent of the dataset. 
Given that the same forward process is applied to these two samples by injecting the same noise, we can express the distance between them as follows:
\begin{equation}
||\mathbf{x}_{T^*} - \mathbf{x}_{T^*}'||_2 = ||\sqrt{\bar{\alpha}_{T^*}} \mathbf{x} - \sqrt{\bar{\alpha}_{T^*}} \mathbf{x}'||_2 \leq ||\mathbf{x} - \mathbf{x}'||_2
\end{equation}

The validity of the aforementioned inequality holds due to the fact that for all $t > 0$, $\bar{\alpha}_t \leq 1$.
\end{proof}

The theorem above convincingly establishes that in a truncated forward process, the $L_2$ distance between the original and compromised PMU data samples exhibits a decreasing trend. 
This reduction in distance indicates the convergence of the compromised PMU data sample towards the original.
Furthermore, the subsequent theorem proves that, throughout the backward process, the trend of decreasing $L_2$ distance continues to persist.

\subsubsection{$L_2$ Distance Analysis during the Backward Process}

The noise estimator $\boldsymbol\epsilon_{\theta}(\mathbf{x}_t, t)$ is trained to predict noise using $\mathbf{x}_t$ and $t$ as inputs. 
The associated loss function is presented in \eqref{eq:loss}. 
By integrating \eqref{eq:forward} with \eqref{eq:loss}, the loss function for $\mathbf{x}_t$ is formulated as follows:

\vspace{-1em}
\begin{equation}
L_\theta(\mathbf{x}_t, t) = ||\frac{\mathbf{x}_t - \sqrt{\bar\alpha_t}\mathbf{x}_0}{\sqrt{1-\bar\alpha_t}} - \boldsymbol\epsilon_{\theta}(\mathbf{x}_t, t)||_2
\label{eq:xtloss}
\end{equation}

The following theorem quantifies how the backward process affects the distance between the original and compromised PMU data samples.

\begin{theorem}
For $\forall t$, if the loss function \eqref{eq:xtloss} satisfies the conditions $||L_\theta(\mathbf{x}_t + \Delta\mathbf{x}_t, t) - L_\theta(\mathbf{x}_t, t)||_2 \leq \frac{1}{2}C_t ||\Delta\mathbf{x}_t||_2$, and $||\Delta\mathbf{x}_t||_2 \leq \xi$, then the $L_2$ distance between the original and compromised PMU data samples continuously decreases throughout the backward process \eqref{eq:backward}. Note that $C_t$ is defined in \eqref{eq:constant}. The theorem can be expressed formally as:
\begin{equation}
||\mathbf{x}_{t-1}' - \mathbf{x}_{t-1}||_2 \leq ||\mathbf{x}_{t}' - \mathbf{x}_{t}||_2
\label{eq:backineq}
\end{equation}
\end{theorem}


\begin{proof}
Assume that the inequality \eqref{eq:backineq} is valid from step $T^*$ to $t$, we will show that it also holds for step $t-1$.
Let $\Delta \mathbf{x}_t = \mathbf{x}_{t}' - \mathbf{x}_{t}$ represent the difference between the original and compromised PMU data sample at step $t$. 
Initially, based on \eqref{eq:adv_attack}, the magnitude of the adversarial attack perturbation satisfies $||\boldsymbol\eta||_2 \leq \xi$.
Furthermore, Theorem \ref{theorem:1} indicates that the forward process reduces the $L_2$ distance between the original and compromised PMU data samples.
Given these premises, it follows that from step $T^*$ to $t$, $||\Delta \mathbf{x}_t||_2 \leq \xi$.

Combining \eqref{eq:xtloss} with the preceding assumption, we can derive the changes in the noise estimator with respect to the changes in the input vector $\mathbf{x}_t$, as illustrated below:
\begin{equation}
\begin{split}
\boldsymbol\epsilon_{\theta}(\mathbf{x}_t + \Delta\mathbf{x}_t, t) - \boldsymbol\epsilon_{\theta}(\mathbf{x}_t, t) =\frac{\Delta\mathbf{x}_t}{\sqrt{1-\bar\alpha_t}} + \boldsymbol\gamma, \\ 
\end{split}
\label{eq:sensitivity}
\end{equation}
where $||\Delta\mathbf{x}_t||_2 \leq \xi \text{ , and } ||\boldsymbol\gamma||_2 \leq C_t ||\Delta\mathbf{x}_t||_2$.

By combining the equation describing the backward process \eqref{eq:backward}, with \eqref{eq:sensitivity} which quantifies the changes of the noise estimator, we can express the PMU data samples' difference at step ($t-1$), as follows:

\vspace{-1em}
\begin{equation}
\begin{split}
\mathbf{x}_{t-1}' - \mathbf{x}_{t-1} & = \frac{1}{\sqrt{{\bar\alpha}_t}}(\mathbf{x}_t' - \mathbf{x}_t) - (\frac{\sqrt{1 - \bar{\alpha}_t}}{\sqrt{{\bar\alpha}_t}} -\sqrt{1 - \bar{\alpha}_{t-1}}) \\
& [\boldsymbol\epsilon_\theta(\mathbf{x}_t + \Delta\mathbf{x}_t, t) - \boldsymbol\epsilon_\theta(\mathbf{x}_t, t)] \\
& = \frac{\sqrt{1 - \bar\alpha_{t-1}}}{\sqrt{1 - \bar\alpha_{t}}} \Delta\mathbf{x}_t + 
(\frac{1}{\sqrt{\bar\alpha_{t}}} - \frac{\sqrt{1 - \bar\alpha_{t-1}}}{\sqrt{1 - \bar\alpha_{t}}})\boldsymbol\gamma
\end{split}
\label{eq:backward_onestep}
\end{equation}

We define the constant $C_{\epsilon_t}$ and $C_t$ as:

\begin{equation}
\begin{split}
C_{\epsilon_t} = (\frac{1}{\sqrt{\bar\alpha_{t}}} - \frac{\sqrt{1 - \bar\alpha_{t-1}}}{\sqrt{1 - \bar\alpha_{t}}}) \text{, }
C_t = (1 - \frac{\sqrt{1 - \bar\alpha_{t-1}}}{\sqrt{1 - \bar\alpha_{t}}}) \frac{1}{C_{\epsilon_t}}
\end{split}
\label{eq:constant}
\end{equation}

Given that for all $t > 0$, $0 < \bar{\alpha}_t < \bar{\alpha}_{t-1} < 1$, it can be shown that $C_{\epsilon_t} > 0$ and $C_t > 0$.
Consequently, the following inequality can be derived from \eqref{eq:backward_onestep}:
\begin{equation}
\begin{split}
||\mathbf{x}_{t-1}' - \mathbf{x}_{t-1}||_2 & \leq \frac{\sqrt{1 - \bar\alpha_{t-1}}}{\sqrt{1 - \bar\alpha_{t}}} ||\Delta \mathbf{x}_t||_2 +   C_{\epsilon_t} C_t ||\Delta \mathbf{x}_t||_2 \\
& = ||\Delta \mathbf{x}_t||_2 = ||\mathbf{x}_{t}' - \mathbf{x}_{t}||_2
\end{split}
\end{equation}
\vspace{-1em}

Assuming the inequality holds true from steps $T^*$ to $t$, it can be inferred that the inequality also holds at step $t-1$. Employing the induction, we can conclude that the inequality maintains its validity in every step of the backward process, given the initial assumption.
\end{proof}

The two theorems discussed earlier illustrate that the $L_2$ distance between the original and compromised PMU data sample consistently decreases during the forward and backward processes of our proposed adversarial purification process. 
This suggests that the impact of perturbations introduced by the adversarial attack diminishes through the entire purification process. The numerical study section below will track the change of the $L_2$ distance in both the forward and backward processes of our proposed adversarial purification algorithm. 
The empirical findings corroborate the theoretical results, demonstrating consistency between experiment and theory.

\section{Numerical Study}

In this section, we conduct a numerical analysis to evaluate the efficacy of the proposed diffusion model-based adversarial purification algorithm in countering adversarial attacks targeting the power system event classifier.
We benchmark the proposed method against multiple baselines and state-of-the-art adversarial purification methods, including Feature squeeze, Singular Value Decomposition (SVD), Low-Pass Filter, and PMU event participation decomposition.

\vspace{-1em}
\subsection{Data Source} \label{DataSource}

The datasets used in this study come from the Phasor Measurement Units (PMUs) of the Western Interconnection in the United States. 
The dataset includes 41 PMUs distributed across diverse geographical regions.
The data pre-processing pipeline involves removing malfunctioning PMUs using either the PMU status or established outlier thresholds. Additionally, this pipeline includes methods for managing missing data and computing real and reactive power.
The raw measurement data consists of voltage and current phasors. 
The data undergoes the pre-processing pipeline detailed in Section III of \cite{cheng2022}. 
The resulting tensor comprises active power ($P$), reactive power ($Q$), voltage magnitude ($|V|$), and frequency ($F$).

System event labels are obtained from the event log maintained by the electric utility and system operators.
The dataset contains 2,468 labeled PMU data samples, classified into four event types: normal system operation behavior (totaling 594), voltage-related events (totaling 624), frequency-related events (totaling 666), and oscillation events (totaling 584).

Each PMU data sample in the dataset represents a 12-second window.
Given that the PMUs operate at a sampling rate of 30 Hz, the shape of each data sample for the Western Interconnection datasets is [360; 41; 4].
These dimensions correspond to the number of timestamps (360), the total number of PMUs (41), and the four measurement channels: active power ($P$), reactive power ($Q$), voltage magnitude ($|V|$), and frequency ($F$).

The datasets are systematically divided into three subsets: training sets, validation sets, and testing sets. 
The training sets encompass 60\% of the total samples, serving as the primary source for model learning. 
Validation sets, representing 20\% of the total samples, are utilized to tune models' hyper-parameters and prevent overfitting. 
The testing sets constitute the remaining 20\%, providing an unbiased evaluation of the final model.
The event classifier and the diffusion model are trained on a machine with a Quadro RTX 6000 GPU.

The method of grid search is employed to select hyper-parameters.
This is conducted over the hyper-parameters: total timestep, $T$, with values $\{20, 50, 100, 200\}$, the truncate timestep, $T^*$, with values $\{0.1 T, 0.2 T, \cdots, 0.9 T\}$, and the backward steps, $S$, with values $\{1, 2, 3\}$.
After evaluation on the validation dataset, the optimal set of hyper-parameters is $T = 20$, $T^*=0.2T$, and $S=3$.

\vspace{-0.5em}
\subsection{Baseline Adversarial Defense Methods}

In this study, after subjecting the dataset to a range of attack algorithms, we adopted multiple baseline adversarial defense algorithms to evaluate the performance of the proposed adversarial purification algorithm.
These purification algorithms include 
Feature Squeezing, Low-pass Filtering, Singular Value Decomposition (SVD), and Event Participation Decomposition. 
A brief description is provided for each of the baseline methods below.


\subsubsection{Feature Squeezing}

Feature Squeezing is an adversarial purification mechanism proposed by Xu et al. \cite{Xu_2018} to mitigate the effects of adversarial attacks.
This method encompasses the exploitable search space of such attacks.
Feature Squeezing encompasses two distinct techniques: 1) bit-depth reduction, which diminishes the bit depth of individual measurements, 2) spatial smoothing, where values are replaced with the average of neighboring measurements.
These two techniques address two perturbation types: small changes on numerous measurements and significant changes on few measurements.

\subsubsection{Low-pass filtering}

The low-pass filter technique is a widely used method in signal processing that effectually retains low-frequency components and filters out high-frequency components.
In the context of power systems, adversarial attacks could involve the injection of malicious perturbations, analogous to high-frequency noise in a signal. 
Consequently, low-pass filtering can serve as an effective purification method against such adversarial attacks. 
In this study, a tenth-order Butterworth low-pass filter is applied to PMU time-series measurements with a cut-off frequency of 10 Hz to mitigate the detrimental impact of adversarial attacks.

\subsubsection{Singular Value Decomposition (SVD) Decomposition}

The Singular Value Decomposition (SVD) is a matrix decomposition technique that factorizes a given matrix into three constituent matrices, possessing notable algebraic properties and providing significant geometric insights into the underlying data structure. 
Utilizing SVD decomposition, the PMU data matrix is reconstructed by selectively preserving the predominant singular values, thus yielding a smoother signal.
This process effectively mitigates adversarial attacks present in the noise space.

\subsubsection{Event Participation Decomposition}

This baseline method is derived from a matrix decomposition technique specifically designed for PMU event data, as proposed by Foggo et al. in \cite{foggo2021phasor}.
This methodology decomposes PMU data samples into the event participation factor, which is designed to encapsulate the features of the event.
Given that adversarial attacks typically induce only minor perturbations to the PMUs, their impact on the event participation factor is negligible, allowing for the reconstructed sample to maintain the integrity of event behavior while mitigating the influence of the attack. 

\vspace{-1em}
\subsection{Performance of Adversarial Purification Methods}

The original and compromised PMU measurements are processed using both the proposed diffusion model-based adversarial purification approach and baseline methodologies. 
The F1 scores of the power system event classifier with the proposed and benchmark adversarial purification methods are systematically summarized in Table \ref{accuracyResult}. 
The first row of the table shows that different adversarial attacks can successfully fool the pre-trained deep neural network-based event classifier successfully. 
Even the least sophisticated adversarial attack algorithm, FGSM, can reduce the F1 score of the event classifier from 96.66\% to 51.3\%. The most powerful adversarial attack algorithm, C\&W attack, can reduce the classifier's F1 score to only 8.5\%, effectively disabling the classifier. 

\begin{table}[ht]
\vspace{-1em}
	\setlength{\tabcolsep}{0.5pt}	
	\caption{F1 Scores of Power System Event Classifier with the Proposed and Baseline Adversarial Purification Methods}
	    \centering
		\begin{tabular}{|c|c|c|c|c|c|c|}
                \hline
                \backslashbox{\textbf{Purifications}}{\textbf{Attacks}} & \textbf{Original} & \textbf{FGSM} & \textbf{PGD} & \textbf{BIM} & \textbf{DeepFool} & \textbf{C\&W} \\ 
                \toprule
                \hline
                No Purification & \textbf{96.7\%} & 51.3\% & 38.1\% & 42.2\% & 51.2\% & 8.5\% \\ 
                \hline
                Feature Squeezing & 95.1\% & 60.5\% & 57.4\% & 86.8\% & 89.1\% & 41.7\% \\
                \hline
                Low-Pass Filtering & 96.4\% & 64.9\% & 60.4\% & 92.7\% & 94.1\% & 42.4\% \\
                \hline
                SVD Decomposition & 95.0\% & 82.6\% & 83.8\% & 92.8\% & 92.9\% & 80.1\% \\ 
                \hline
                Event Decomposition & \textbf{96.7\%} & 63.2\% & 58.0\% & 94.4\% & \textbf{96.2\%} & 38.2\% \\ 
                \hline
                Proposed Method & 96.1\% & \textbf{92.7\%} & \textbf{92.8\%} & \textbf{94.9\%} & 95.0\% & \textbf{91.4\%} \\ 
                \hline
		\end{tabular}%
	\label{accuracyResult}%
\vspace{-0.5em}
\end{table} 

The feature squeezing, low-pass filtering, and event decomposition methods can purify some of the compromised PMU data samples in the dataset. However, the F1 score of the classifier performance is still below 50\% when the data is subject to C\&W attack. 
Among all the baseline methods, the SVD is the most effective method, purifying over 80\% of the compromised PMU data samples subject to all types of adversarial attacks.
For the BIM and DeepFool attack, the SVD method helps improve the classifier's F1 score on the purified PMU measurements to over 90\%. However, for the FGSM and PGD attack, even with the support of SVD, the classifier's F1 score can only reach around 82\%.

Our proposed diffusion model-based adversarial purification method is much more effective than all other baseline methods. Facilitated by our proposed method, power system event classifier could reach F1 scores of over 91\% under all attack scenarios. It achieves the highest F1 score when the PMU data is compromised by the FGSM, PGD, BIM and C\&W attacks. In the C\&W and FGSM attack scenarios, the proposed algorithm increases the F1 score from the state-of-the-art baseline algorithm by more than 11\%.
In sum, the proposed diffusion model-based adversarial purification technique demonstrates superior performance in comparison to all baseline procedures across the full spectrum of adversarial attacks on large-scale real-world PMU datasets. 

\begin{figure}[!t]
\centering
\includegraphics[width=0.48\textwidth]{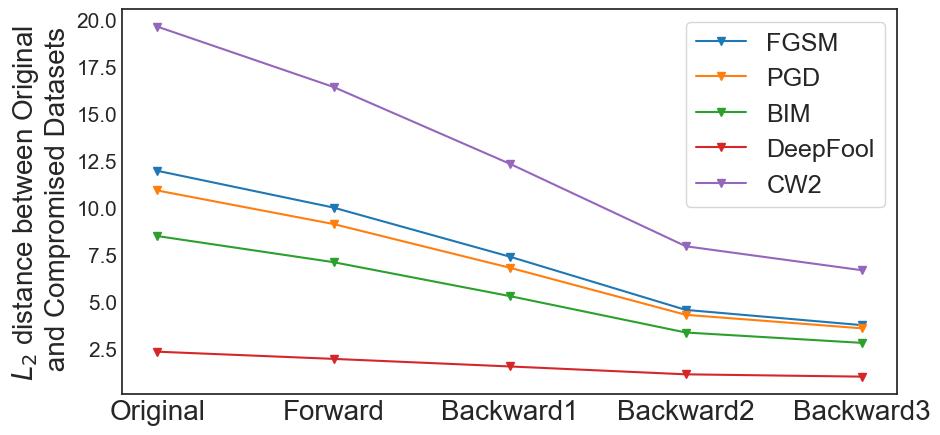}
\DeclareGraphicsExtensions.
\vspace{-1.5em}
\caption{$L_2$ distance between the adversarial and original PMU datasets during the forward and backward processes of the adversarial purification method.}
\label{l2distance_ddim}
\vspace{-1.5em}
\end{figure}

\vspace{-1em}
\subsection{Empirical Analysis of $L_2$ Distance of the Diffusion Model-based Adversarial Purification Method}

Figure \ref{l2distance_ddim} depicts the average $L_2$ distance between the original PMU dataset and the compromised PMU dataset during the forward and backward phases of our proposed adversarial purification algorithm. 
Upon close examination of the figure, two conclusions can be drawn. 
First, a decrease in the $L_2$ distance is observed during the forward process, a finding that aligns with the result from Theorem 1. 
Second, a similar reduction in the $L_2$ distance is noted during the backward process, which corroborates Theorem 2. 
It is of importance to note that the initial phase of the backward process is mainly responsible for the substantial decrease in the $L_2$ distance between the two datasets. 
Thus, the effectiveness of our proposed adversarial purification method can be substantiated, not only through theoretical premises but also through empirical evidence derived from the real-world PMU dataset.

\vspace{-1em}
\subsection{Computation Time of Adversarial Purification Methods}

In order to provide real-time monitoring for power systems, existing PMUs typically adopt a 30Hz sampling rate, underlining the importance of computation efficiency for adversarial purification methods for PMU data.
A comparative analysis is conducted to quantify the computation time of different adversarial purification methods. 
We conduct simulations with streaming PMU data and apply adversarial purification algorithms to each incoming PMU measurement. Subsequently, we calculate the average processing time required for these measurements.
To ensure a fair comparison, we leveraged multi-threading and GPU acceleration to enhance the performance of different algorithms. 
For the feature squeezing and low-pass filtering methods, we employ separate threads to compute data from individual PMUs and to measure the average processing time. For the SVD and event participant decomposition methods, each measurement channel's data matrix is computed using distinct threads. The proposed diffusion-based method is processed and measured on the GPU.
To investigate the scalability of various methods, simulations were conducted using varying numbers of PMUs.
The results are presented in the table below.

\begin{table}[ht]
\setlength{\tabcolsep}{3.5pt}	
\vspace{-1em}
\caption{Computation time of the baseline and proposed adversarial purification methods for PMU Data.}
\vspace{-0.5em}
	    \centering
		\begin{tabular}{|c|c|c|c|c|}
		    \hline
                \multirow{2}{*}{\backslashbox{\textbf{Purifications}}{\textbf{PMU Number}}} & 40 & 80 & 120 & 160 \\ 
                \cline{2-5} & \multicolumn{4}{c|}{Average Measurement Process Time} \\
                \hline
                Feature Squeezing & 1 ms & 2 ms & 3 ms & 4 ms \\
                \hline
                Low Pass Filtering & $<$1 ms & 1 ms & 2 ms & 3 ms \\
                \hline
                SVD Decomposition & 12 ms & 27 ms & 45 ms & 76 ms \\ 
                \hline
                Event Decomposition & 37 ms & 85 ms & 130 ms & 178 ms \\ 
                \hline
                Proposed Method & 18 ms & 24 ms & 29 ms & 34 ms \\ 
                \hline
            \end{tabular}%
\label{VoltageEventResult}%
\vspace{-0.5em}
\end{table} 

As shown in Table \ref{VoltageEventResult}, the low-pass filtering method has the highest computation efficiency as the method only processes a few neighboring measurements with filter coefficients to purify the incoming PMU measurements. 
Moreover, as the count of PMUs escalates, the time consumed increases in a linear manner. 
The feature-squeezing technique operates through a dual-step process: bit-depth reduction and spatial smoothing, both following a linear time complexity.
Consequently, the computational efficiency of the feature-squeezing method is comparable to that of the low-pass filtering technique. However, due to its two-step process, the feature-squeezing method is marginally slower.

Both the SVD and event participation decomposition methods employ matrix decomposition techniques.
Given a channel's PMU measurement matrix of dimensions $W \times k$, where $W$  represents the time steps and $k$ indicates the number of PMUs, the time complexity for both methods is $O(W^2k + k^3)$.
In our experiment, the time step $W$ is kept constant and is considerably larger than the number of PMUs. 
As a result, the first term ($W^2k$) significantly influences the overall computational complexity. 
This implies that the processing time grows linearly as the number of PMUs increases, aligning with the findings in Table \ref{VoltageEventResult}.
Nevertheless, if the number of PMUs continues to grow, the cubic term in the time complexity will dominate, leading to substantial computational demands.

Our proposed method, based on a diffusion model, comprises two steps: the truncated forward and backward processes. 
The forward process, described by \eqref{eq:forward}, exhibits linear time complexity relative to the number of PMUs. 
The backward process, given by \eqref{eq:backward}, involves $S$ times neural network inferences.
Since the inference layers within the neural network are processed in parallel by the GPU's CUDA cores, increasing the number of PMUs does not notably increase the computation time.
As evidenced in Table \ref{VoltageEventResult}, our method displays relatively consistent computational time with an increasing number of PMUs, offering enhanced scalability compared to matrix decomposition-based methods.

\section{Conclusion}

The growing threat of cyber-attacks, particularly adversarial attacks, has prompted researchers in the field of power system to develop effective defense strategies which focus on protecting machine learning-based models.
This paper proposes an innovative diffusion model-based adversarial purification algorithm designed to defend against adversarial attacks on power system event classifiers. 
The algorithm comprises two components: a truncated forward process, which injects noise into the data, and a denoising backward process facilitated by a trained noise estimator.
The numerical study with a large-scale real-world PMUs dataset demonstrates that our algorithm outperforms the state-of-the-art adversarial purification methods in classification accuracy under various adversarial attacks. By incorporating DDIMs to omit intermediate steps in the backward process, our algorithm achieves remarkable computational efficiency, meeting the demands of real-time operation. Moreover, a theoretical analysis of the $L_2$ distance throughout the adversarial purification process reveals that our method continuously reduces the $L_2$ distance between the original and compromised PMU data, shedding light on the working mechanism of the proposed technique. 
\bibliographystyle{IEEEtran}
\bibliography{IEEEabrv,ref}

\end{document}